\documentstyle[twocolumn,seceq]{jpsj}

\input epsf  

\def\runtitle{Symmetry of high-piezoelectric Pb based complex perovskites
 at the morphotropic . . . .}
\def\runauthor{Yasusada {\sc Yamada} $et\ al$.}

\title
{
Symmetry of high-piezoelectric Pb based complex perovskites at the 
morphotropic phase boundary II. Theoretical treatment
}

\author
{ 
Yasusada {\sc Yamada}$^{1,2,}$\footnote{E-mail: yasusada@mn.waseda.ac.jp}, 
Yoshiaki {\sc Uesu}$^{1,3}$, Masaaki {\sc Matsuda}$^{2}$, 
Kouji {\sc Fujishiro}$^{4}$,
 Dave E. {\sc Cox}$^{5}$, Beatriz {\sc Noheda}$^{5}$
  and Gen {\sc Shirane}$^{5}$
}

\inst
{
$^1$Advanced Research Institute, Waseda University, 
3-4-1 Okubo, Shinjuku-ku, Tokyo 169- 8555\\
$^2$Advanced Science Research Center, Japan Atomic Energy 
Research Institute, Tokai, Ibaraki 319- 1195  \\
$^3$  Department of Physics, Waseda University, 3-4-1 
Okubo, Shinjuku-ku, Tokyo 169- 8555\\
$^4$  Materials and Structures Laboratory, Tokyo Institute 
of Technology,	4259 Nagatsuda, Midori-ku, Yokohama 226- 8503\\
$^5$  Department of Physics, Brookhaven National Laboratory, 
Upton, New York 11973\\
}

\recdate
{
\today
}

\abst
{
The structural characteristics of the perovskite-based ferroelectric 
$\mathrm{Pb(Zn_{1/3}Nb_{2/3})_{1-x}Ti_{x}O_{3}}$  at the morphotropic 
phase boundary (MPB) region (x$\simeq$ 0.09) have been analyzed. 
The analysis is based on the symmetry adapted free energy functions 
under the assumption that the total polarization and the unit cell 
volume are conserved during the transformations between various 
morphotropic phases. Overall features of the relationships between 
the observed lattice constants at various conditions have been 
consistently explained. The origin of the anomalous physical 
properties at MPB is discussed.
}

\kword
{
Piezoelectricity, morthotropic phase boundary, neutron diffraction,
$\mathrm{ Pb(Zn_{1/3}Nb_{2/3})O_{3} / 9\% PbTiO_{3}}$
}

\begin{document}
\sloppy
\maketitle

\section{Introduction}

The physical properties in solids at the phase boundaries between the 
two competitive stable phases tend to exhibit anomalous characteristics. 
A typical example is seen in the extraordinary transport phenomena such 
as high-$\mathrm{T_{c}}$  superconductivity and colossal magnetoresistance 
at metal-insulator phase boundary of the transition metal oxides. 
The high piezoelectric effect at the morphotropic phase boundary 
(MPB) observed in several perovskite-based ferroelectric materials 
may be considered to provide another interesting example of such 
phenomena at the structural phase boundary.
\begin{figure}[htbp]
\centering{ \epsfxsize 7cm \epsfbox{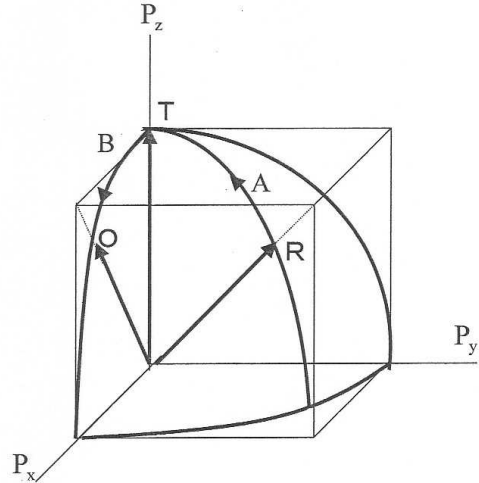}}
\caption[Fig.1]{
The trajectory followed by $\mbox{\boldmath$P$}_s$ in the 3-dimensional 
polarization vector space. R, T and O represent the symmetries of the 
structure as $\mbox{\boldmath$P$}_s$ is oriented to [111], [001], and [101] 
respectively. The symbols A and B indicate the process between R$\rightarrow$T 
and T$\rightarrow$R respectively. 
}
\label{fig:1}
\end{figure}

As is discussed in Part I\cite{rf:1} of the present paper, the high 
piezoelectricity has been investigated from both experimental as well 
as theoretical aspects. Experimentally, Noheda $et\ al$.\cite{rf:2,rf:3,
rf:4,rf:5} discovered a previously unknown 
monoclinic phase belonging to space 
group $\mathrm{Cm}$ ($\mathrm{M_{a}}$-phase) in the boundary region between 
rhombohedral (R) and tetragonal phases in 
$\mathrm{Pb_{1}Zr_{1-x}Ti_{x}O_{3}}$ (abbreviated as PZTx). This is 
immediately followed by the theoretical work by Bellaiche $et\ al$..\cite{rf:6}  
Based on the first principles calculations they showed that there should 
in fact exist $\mathrm{M_{a}}$-phase within a narrow concentration 
range intervening between the high symmetry R- and T- phase in PZTx 
system in complete agreement with the experimental results. It is also 
pointed out that the high piezoelectricity is associated with the 
stabilization of the monoclinic phase.

Subsequently, X-ray measurements by Noheda $et\ al$.\cite{rf:7} on 
another type of material 
$\mathrm{Pb(Zn_{1/3}Nb_{2/3})_{1-x}Ti_{x}O_{3}}$
( abbreviated as PZN-xPT ) also revealed the existence of similar 
monoclinic phase at the MPB region, except that the space group in 
this case is $\mathrm{Pm}$ ($\mathrm{M_{c}}$-phase) instead 
of $\mathrm{Cm}$.

On the other hand,
Ishibashi and Iwata\cite{rf:m1,rf:m2,rf:m3} developed a phenomenological 
theory on the physical properties at morphotropic phase boundary in 
perovskite solid solutions. He pointed out that depending on the 
coefficients of fouth order terms of the free energy expansion, the
system stabilizes the orthorhombic phase intervening between the
tetragonal and the rhombohedral phases. He also discussed the
dielectric\cite{rf:m1,rf:m2} as well as elastic\cite{rf:m3} responses of the 
system at MPB and pointed out that the anomalous behavior in the 
susceptibility tensor components is in fact due to large electromechanical
coupling constants of these materials.
Fu and Cohen\cite{rf:8} presented the key concept to 
understand the physics of high piezoelectricity: that is, the 
rotation of the spontaneous polarization during the structural 
transitions under electric field. Through first principles calculations 
they showed that there is a low energy path along the lines combining 
the symmetric directions of [111], [001] and [101] as shown in Fig.1. 
Upon application of the external field, the polarization would easily 
rotate along the particular path, which results in a large piezoelectric 
response. The stabilization of the monoclinic phase are considered to 
manifest this view point since monoclinicity is induced during the 
polarization vector continuously rotates along the specific path.

As is discussed by Noheda $et\ al.$\cite{rf:7}, in the case of PZN-8PT, 
the $\mathrm{M_{c}}$-phase appears on the path between 
$\mbox{\boldmath$P$}$ $\parallel$ [001] and $\mbox{\boldmath$P$}$ 
$\parallel$ [101]. 
This suggests that the high symmetry stable phase 
for composition around 8PT 
would be tetragonal and orthorhombic (O) rather than rhombohedral. 
In fact, depending on the prehistory of the sample preparation such 
as heat treatment, poling, etc., the orthorhombic phase is 
observed to become stabilized in PZN-9PT powder sample\cite{rf:9}.

Recently, we carried out a neutron diffraction study using single 
crystals of PZN-9PT, with focus on the existence of $\mathrm{M_{c}}$-phase. 
The results are summarized in Part I of the present paper. In the present 
study, we develop a theoretical treatment based on a phenomenological 
arguments in order to analyze the experimental results presented in Part I. 
Since the treatment is phenomenological, unlike the treatment based on 
the first principles calculations, the conclusions depend on a few 
disposable parameters included in the free energy expression. 
Nevertheless, the results seem to provide some important insights 
into the specific physics at MPB regions.

\section{Analysis of Experimental Data}

\begin{fullfigure}[htbp]
\centering{ \epsfxsize 14cm \epsfbox{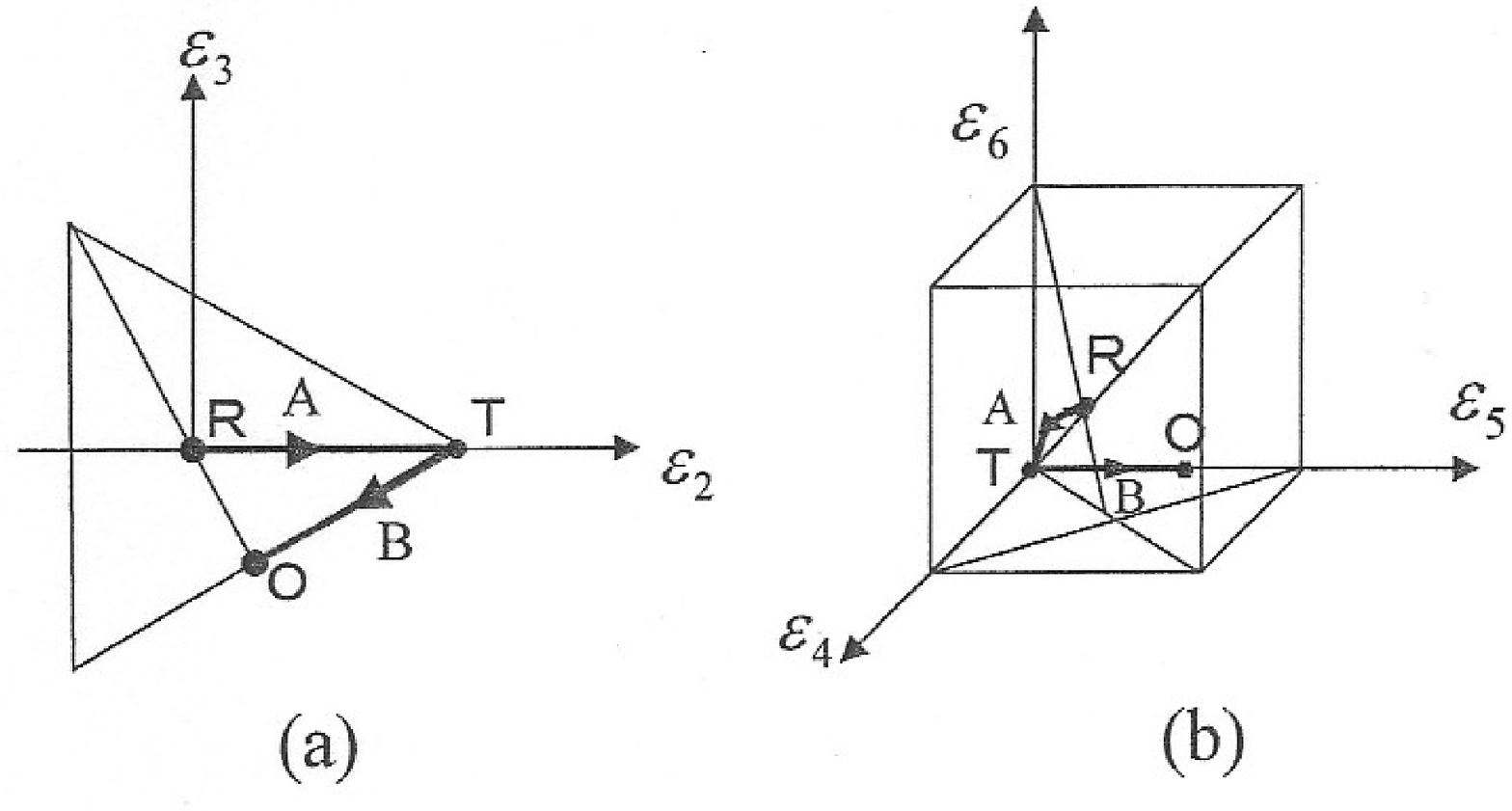}}
\caption[Fig.2]{
(a)	The trajectory followed by the lattice distortions in 
the 2-dimensional space spanned by the basis of $E_{g}$ irreducible 
representation.
(b)	The trajectory followed by the lattice distortions in the 3-dimensional 
space spanned by the basis of $T_{2g}$ irreducible representation.
}
\label{fig:2}
\end{fullfigure}
Using $\mathrm{BaTiO_{3}}$ as a model system, Fu and Cohen\cite{rf:8} 
developed a detailed theoretical treatment and discussed that the 
morphotropic phase transition scheme in $\mathrm{BaTiO_{3}}$ is 
described by the 'rotation' of the spontaneous polarization vector, 
$\mbox{\boldmath$P$}_s$ , within the $(P_{x}, P_{y}, P_{z})$ space moving 
along [111]$\rightarrow$[001]$\rightarrow$[101] directions 
as shown in Fig. 1. This process 
realizes the phases with rhombohedral, tetragonal and orthorhombic 
symmetry. Later, Vanderbilt and Cohen\cite{rf:10} pointed out that when 
the energy terms up to the eighth order with respect to polarization 
are included in the free energy expansion, the monoclinic phases with 
the polarization along [$\xi \ \xi \ \zeta$]-($\mathrm{M_{a}}$-phase) 
and [$\xi \ 0 \ \zeta$]-($\mathrm{M_{c}}$-phase) directions can be stabilized.

Following these previous arguments, we also assume that the 
polarization moves on the surface of the sphere 
with $|\mbox{\boldmath$P$}_{s}|$ 
conserved during the process of the polymorphic phase transitions in 
the present system. For later convenience, let us define the parameters 
to specify the orientation of polarization as it moves along the 
trajectories A and B in the $\mbox{\boldmath$P$}$-space (See Fig.1.) 
as follows:

process A:
\begin{eqnarray}
\rho=\frac{P_{z}^{2}}{P^2}\  , \label{eq:1}
\end{eqnarray}

process B:
\begin{eqnarray}
\rho'=\frac{P_{x}^{2}}{P^2}\ , \label{eq:2}
\end{eqnarray}
In the process A, $\rho$ changes in the range 
of $\frac{1}{3}\leq\rho\leq1$ as $\mbox{\boldmath$P$}$ rotates from 
[111] $(\rho=\frac{1}{3})$ to [001] ($\rho$=1), while in the 
process B, $\rho'$ changes in the range of 
$0\leq\rho'\leq\frac{1}{2}$ as $\mbox{\boldmath$P$}$ rotates from 
[001] ($\rho'$=0) to [101] $(\rho'=\frac{1}{2})$.

In addition, we further assume that the volume of the unit cell is 
also conserved throughout the sequential phase transitions. As is 
typically seen in the case of $\mathrm{BaTiO_{3}}$, this assumption seems to 
be satisfied well.

Since we are particularly interested in the lattice distortions, 
we give the expression of the 'elastic' free energy leaving the 
strain components as the independent variables rather than projecting 
on the polarization space. In order to make the treatment physically 
transparent, we construct the basis functions of the six dimensional 
direct product space 
of $(P_{x}^{2},P_{y}^{2},P_{z}^{2},P_{y}P_{z},P_{z}P_{x},P_{x}P_{y})$ 
and $(e_{xx},e_{yy},e_{zz},e_{yz},e_{zx},e_{xy})$ associated with 
the irreducible representations of the point group m$\bar{3}$m as follows.

\begin{eqnarray}
\psi_{1}=\frac{1}{\sqrt{3}}(P_{x}^{2}+P_{y}^{2}+P_{z}^{2})\in A_{1g} 
\ \ \ \  \nonumber \\
\left. \begin{array}{l}
\psi_{2}=\frac{1}{\sqrt{6}}(2P_{z}^{2}-P_{x}^{2}-P_{y}^{2})  \\
\psi_{3}=\frac{1}{\sqrt{2}}(P_{x}^{2}-P_{y}^{2})
\end{array} \right\} \in E_{g}  \nonumber \\
\left. \begin{array}{l}
\psi_{4}=P_{y}P_{z}  \\
\psi_{5}=P_{z}P_{x}  \\
\psi_{6}=P_{x}P_{y} 
\end{array} \right\} \in T_{2g} \ \ \ \ \ \ \ \ \ \ \ \ \ \ \ \ \ \
\label{eq:3}
\end{eqnarray}
and
\begin{eqnarray}
\ e_{1}=\frac{1}{\sqrt{3}}(e_{xx}+e_{yy}+e_{zz}) \in A_{1g} 
\ \ \ \ \nonumber \\
\left. \begin{array}{l}
\ e_{2}=\frac{1}{\sqrt{6}}(2e_{zz}-e_{xx}-e_{yy})  \\
\ e_{3}=\frac{1}{\sqrt{2}}(e_{xx}-e_{yy}) 
\end{array} \right\} \in E_{g}  \nonumber \\
\left. \begin{array}{l}
\ e_{4}=e_{yz}  \\
\ e_{5}=e_{zx}  \\
\ e_{6}=e_{xy} 
\end{array} \right\} \in T_{2g}  
\ \ \ \ \ \ \ \ \ \ \ \ \ \ \ \ \ \ \ \ \ \ 
\label{eq:4}
\end{eqnarray}                        
Then, we can set up the simplest symmetry adapted elastic free energy including only the quadratic (or bilinear) terms as follows: 

\begin{eqnarray}
\ F_{el}=F_{el}^{(A_{1g})} + F_{el}^{(E_{g})} + F_{el}^{(T_{2g})},
\ \ \ \ \ \ \ \ \ \ \ \label{eq:5}
\end{eqnarray}                       
where
\begin{eqnarray}
\ F_{el}^{(A_{1g})} = \frac{1}{2}ce_{1}^{2} - ge_{1}\psi_{1},
\ \ \ \ \ \ \ \ \ \ \ \ \ \ \ \ \ \ \ \ \ \label{eq:6}
\end{eqnarray}

\begin{eqnarray}
\ F_{el}^{(E_{g})} = \frac{1}{2}c'(e_{2}^{2}+e_{3}^{2}) - g'(e_{2}\psi_{2}+e_{3}\psi_{3}), \label{eq:7}
\end{eqnarray}

\begin{eqnarray}
\ F_{el}^{(T_{2g})} = \frac{1}{2}c''(e_{4}^{2}+e_{5}^{2}+e_{6}^{2}) \ \ \ \ \  
\ \ \ \ \ \ \  \ \ \ \ \ \ \ \nonumber \\
-g''(e_{4}\psi_{4}+e_{5}\psi_{5}+e_{6}\psi_{6}). \label{eq:8}
\end{eqnarray}
The second terms of these equations give the electrostrictive 
energies in the cubic perovskite system.

In these expressions, the independent thermodynamical variables are 
$e_{v}'$s $(v=1,\cdots 6)$ while $\psi_{v}'$s are considered to be 
the parameters to define the orientation of the polarization which are 
explicitly given by $\rho$ and $\rho'$ in eqs. (\ref{eq:1}) and (\ref{eq:2}). 
Among the 
three terms in eq. (\ref{eq:5}), $F_{el}^{(A_{1g})}$  is 
irrelevant to determine 
the stability of the system during the rotation of the polarization 
vector, since it is expressed in terms of $e_{1}$ and $\psi_{1}$ 
which are both assumed to be conserved.
\begin{fulltable}
\caption{The lattice distortions of the perovskite system in 
the polymorphic phases realized when $\mbox{\boldmath$P$}_s$ rotates 
through the process A and B. The disposable parameters are 
defined by: $\Delta=\frac{2g'}{3c'}, \delta=\frac{g''}{3c''}$.}
\label{Table I}
\begin{fulltabular}{p{1cm} ||c c |p{2cm}|p{2cm}|p{2cm}|p{2cm}|p{2cm}|p{2cm}}
\cline{1-9}
   &  & &  $\varepsilon_{xx}$  & $\varepsilon_{yy}$ & $\varepsilon_{zz}$ & 
   $\varepsilon_{yz}$ & $\varepsilon_{zx}$ & $\varepsilon_{xy}$ 
\\ \cline{2-9}
   & $\rho$ & $\rho'$  &  $\Delta a$  & $\Delta b$ & $\Delta c$ & 
   $\Delta \alpha$ & $\Delta \beta$ & $\Delta \gamma$ 
\\ \cline{1-9} 
  R & $\frac{1}{3}$ &  & 0  & 0  & 0 & $\delta$ & $\delta$ & $\delta$ 
\\   \cline{1-9}
  $\mathrm{M_a}$ & $\rho$ & - & $-\frac{(3\rho-1)\Delta}{4}$  
  & $-\frac{(3\rho-1)\Delta}{4}$  & $+\frac{(3\rho-1)\Delta}{2}$ 
  & $3\delta\sqrt{\frac{\rho(1-\rho)}{2}}$ 
  & $3\delta\sqrt{\frac{\rho(1-\rho)}{2}}$ 
  & $3\delta\frac{(1-\rho)}{2}$ 
\\   \cline{1-9}
  T & 1 & 0 & $-\frac{\Delta}{2}$  & $-\frac{\Delta}{2}$  & $\Delta$ 
  & 0 & 0 & 0 
\\   \cline{1-9}
  $\mathrm{M_c}$ & - & $\rho'$ & $\frac{(-1+3\rho')\Delta}{2}$  
  & $-\frac{\Delta}{2}$  & $\frac{(2-3\rho')\Delta}{2}$ 
  & 0 & $3\delta \sqrt{\rho' (1-\rho')}$ & 0 
\\   \cline{1-9}
  O & - & $\frac{1}{2}$ & $\frac{\Delta}{4}$  
  & $-\frac{\Delta}{2}$  & $\frac{\Delta}{4}$ 
  & 0 & $\frac{3}{2}\delta$ & 0 
\\   \cline{1-9}
\end{fulltabular}
\end{fulltable}

Moreover, $F_{el}^{(E_{g})}$ is defined in the 2-dimensional space 
spanned by the basis of $E_{g}$ representation ($\epsilon_{2}$ and 
$\epsilon_{3}$), while $F_{el}^{(T_{2g})}$  in the 3-dimensional 
space spanned by the basis of $T_{2g}$ representation 
($\epsilon_{4},\epsilon_{5}$,and $\epsilon_{6}$). Therefore, 
in order to find the stable lattice distortions, we simply 
minimize $F_{el}^{(E_{g})}$ and $F_{el}^{(T_{2g})}$ 
independently.\\

\noindent (i)$E_g$-space\\

By the standard minimization procedure, it is easily shown that the trajectory of the minimum energy $\epsilon^{0}(\rho)$ is given by:

(a)	Process A
\begin{eqnarray}
\left. \begin{array}{l}
\epsilon_{2}^{0}(\rho) = \frac{g'}{\sqrt{6}c'} (3\rho-1),  \\
\epsilon_{3}^{0}=0.
\end{array} \right\}   \label{eq:9}
\end{eqnarray}                                                      

(b)	Process B
\begin{eqnarray}
\left. \begin{array}{l}
\hat{\epsilon}_{2}^{0}(\rho) = \frac{g'}{\sqrt{2}c'}(1-2\rho'),  \\
\hat{\epsilon}_{3}^{0} = \frac{-2g'}{\sqrt{6}c'}.
\end{array} \right\}   \label{eq:10}
\end{eqnarray} 
where $\hat{\epsilon}_{2}$ and $\hat{\epsilon}_{3}$ are defined by 
120$^{\circ}$ rotation of the ($\epsilon_{2}$,$\epsilon_{3}$) 
coordinate system.

The trajectory followed by the representative point of the system in 
the ($\epsilon_{2}$,$\epsilon_{3}$) space is given in Fig.2(a).
In the figure, R, T and O represents the rhombohedral, tetragonal 
and orthorhombic phases respectively.\\

\noindent (ii)$T_{2g}$-space\\

\noindent Similarly, we have the trajectory of the energy minimum path:

(a)	Process A
\begin{eqnarray}
\left. \begin{array}{l}
\epsilon_{4}^{0}(\rho) = \epsilon_{5}^{0}(\rho) = \frac{g''}{c''} 
\sqrt{\frac{\rho(1-\rho)}{2}}, \\
\epsilon_{6}^{0}(\rho) = \frac{g"}{c"} \frac{1-\rho}{2}.
\end{array} \right\}   \label{eq:11}
\end{eqnarray}                                             

(b)	Process B
\begin{eqnarray}
\left. \begin{array}{l}
\epsilon_{4}^{0} = \epsilon_{6}^{0} = 0,  \\
\epsilon_{5}^{0}(\rho') = \frac{g''}{c''} \sqrt{\rho'(1-\rho')}.
\end{array} \right\}   \label{eq:12}
\end{eqnarray}

The trajectories followed by the representative point of the system 
in the $(\epsilon_{4},\epsilon_{5},\epsilon_{6})$ space 
is given in Fig.2(b).

Eqs. (\ref{eq:9}) to (\ref{eq:12}), together with the condition of the volume 
conservation $(\epsilon_{xx}^{0}+\epsilon_{yy}^{0}+\epsilon_{zz}^{0}=0)$, 
give the set of lattice parameters of the stable state in terms 
of $\rho$ or $\rho'$ as shown in Table I. Notice among total of 30 
lattice constants, there are only two disposable parameters, $\triangle$ 
and $\delta$, besides $\rho$ and $\rho'$. We utilized the values of 
$c_{1}$ (the unique axis in T-phase) observed in 9PT5 and $\beta$ 
(monoclinic angle in O-phase) in 9PT4 as tabulated in Part I 
(see Table I) to obtain the value of $\triangle$ and $\delta$ respectively.

Fig.3 shows the sequence of the calculated lattice constants to be 
taken by the system as $\rho$ and $\rho'$ is varied to stabilize 
(R)$\rightarrow$$\mathrm{(M_{a})}$$\rightarrow$T$\rightarrow$$\mathrm{M_{c}}$$\rightarrow$O, including the 
fictitious stable phases R and $\mathrm{M_{a}}$ which are not 
realized in the PZN-9PT system, but realized in the related materials. 
The observed values at lower temperatures given in Part I are indicated 
by the solid circles. We can not calculate the lattice parameters as a 
function of temperature for direct comparison with the experimental 
results, since the temperature dependences of $\rho (T)$ 
and $\rho'(T)$  are outside of the framework of 
the present treatment.
\begin{figure}[htbp]
\centering{ \epsfxsize 7cm \epsfbox{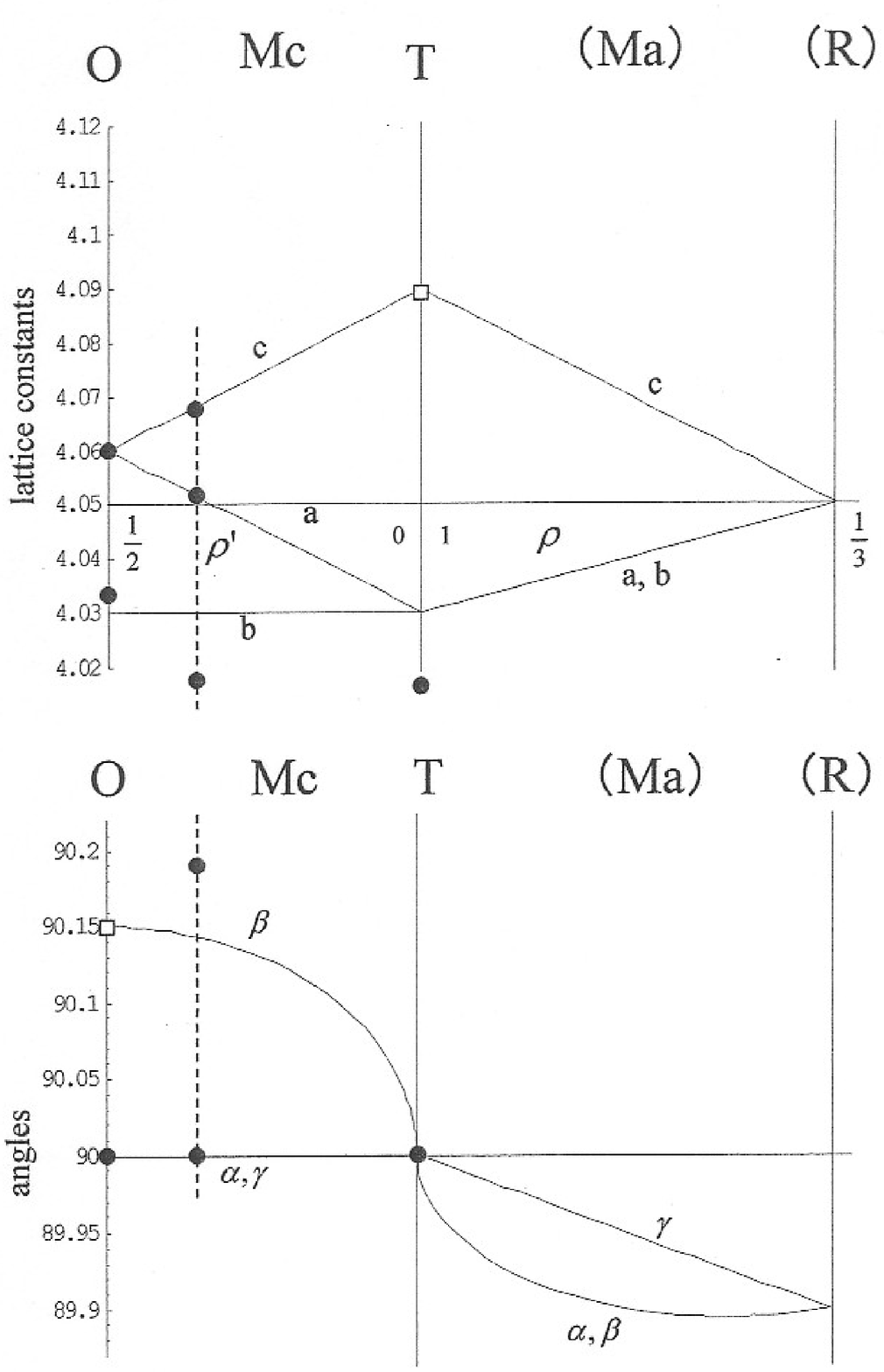}}
\caption[Fig.3]{
The calculated variation of the lattice constants in PZN-9PT as 
the parameters $\rho$ and $\rho'$ is varied. The solid circles are the 
experimental values obtained by the present work (Part I). The figure 
includes the fictitious phases (R and $\mathrm{M_{a}}$) which are not 
realized in 9PT, but realized in the closely related systems. The open 
squares indicate the data used for the numerical analysis. (See the Text.)
}
\label{fig:3}
\end{figure}

\section{Conclusions and Discussions}

In conclusion, based on the symmetry adapted free energy function, 
we analyze the structural characteristics of PZN-9PT under the 
assumption that the total polarization and the unit cell volume 
are conserved during the transformations between various morphotropic 
phases. Overall features of the relationship between the observed 
lattice constants in various phases have 
been consistently explained.

So far, we have discussed the lattice distortions which are 
directly observed by the diffraction studies described in Part I. 
The energetic considerations concerning the stability of the phases 
seem to give more important insight into the physical properties of 
the system at MPB as discussed in the following.

Using eqs. (\ref{eq:9}) through (\ref{eq:12}), the free energy values of the 
rhombohedral $(F_{el}\mathrm{(R)})$, the tetragonal-$(F_{el}\mathrm{(T)})$, 
and the orthorhombic-$(F_{el}\mathrm{(O)})$ phases are given by,

\begin{eqnarray}
\left. \begin{array}{l}
\ F_{el}(R) = -\frac{g''^{2}}{6c''},  \\
\ F_{el}(T) = -\frac{g'^{2}}{3c'},  \\
\ F_{el}(O) = -\frac{g'^{2}}{12c'} - \frac{g''^{2}}{8c''}, 
\end{array} \right\}   \label{eq:13}
\end{eqnarray} 
which means that $(F_{el}\mathrm{(O)})$ is expressed by a fixed weighted 
mean of $(F_{el}\mathrm{(R)})$ and $(F_{el}\mathrm{(T)})$ as,

\begin{eqnarray}
\ F_{el}(\mbox{O}) = \frac{1}{4} \ F_{el}(\mbox{T}) 
+ \frac{3}{4} \ F_{el}(\mbox{R}), \label{eq:14}
\end{eqnarray}
irrespective of the parameter values of $c'$s and $g'$s. Assuming, for 
simplicity, linear dependence of the parameters on the $x$-value 
(the concentration of $\mathrm{PbTiO_{3}}$), the relative energies 
of the three phases change with respect to $x$ as schematically shown 
in Fig.4. We see that the stable phases are generally given by 
either T- or R- phases while O-phase barely becomes one of the 
three degenerated ground states at the single critical 
concentration, $x_{c}$,\\ 
where,
\begin{eqnarray}
\frac{g'^{2}}{3c'} = \frac{g''^{2}}{6c''} \equiv A  \label{eq:15}
\end{eqnarray}
is satisfied. Notice the critical concentration $x=x_{c}$ is nothing but 
the MPB point. (See Fig.4.)
\begin{figure}[htbp]
\centering{ \epsfxsize 7cm \epsfbox{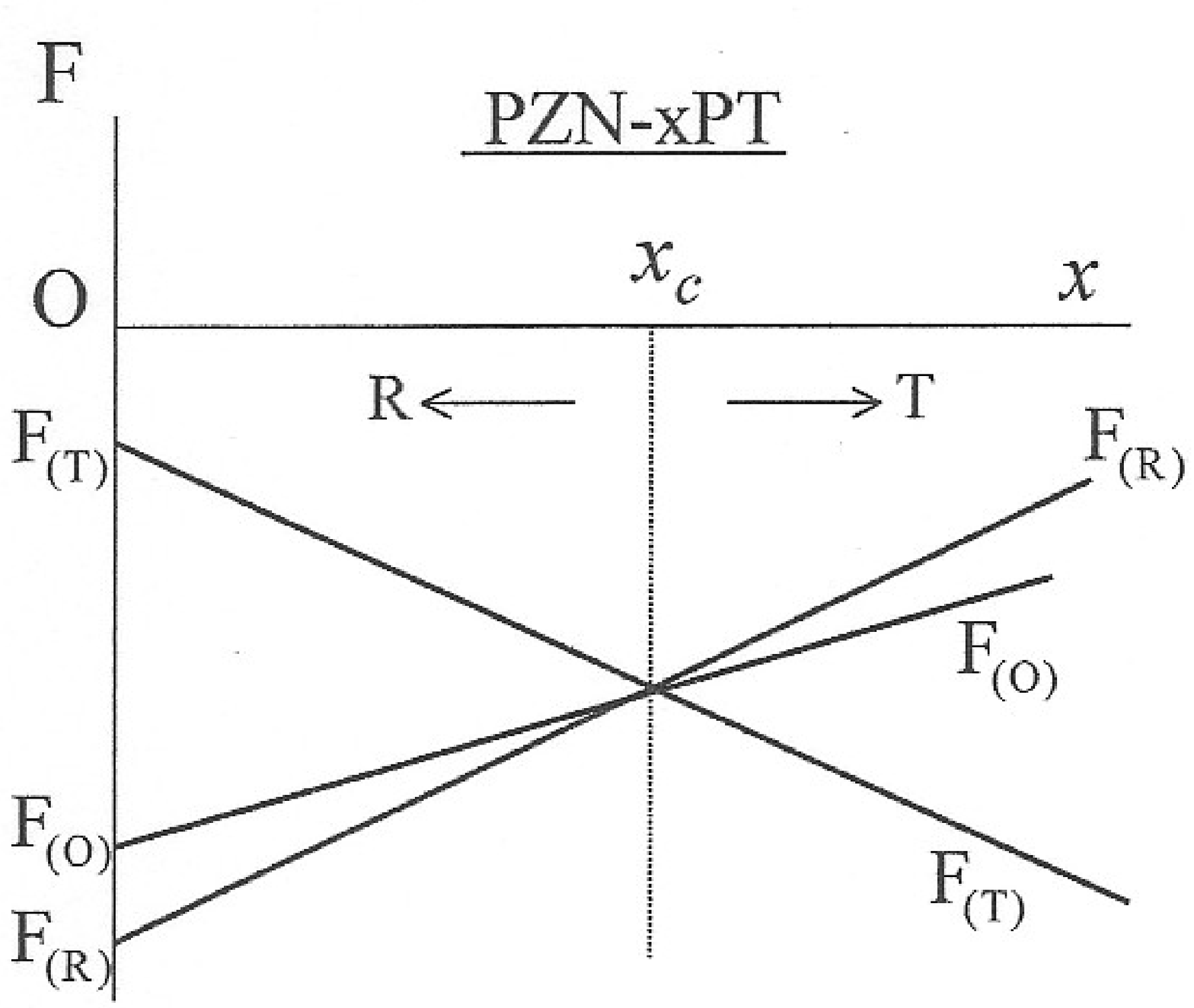}}
\caption[Fig.4]{
Schematic diagram of the relative phase stability as $x$-value 
(the concentration of $\mathrm{PbTiO_{3}}$) is changed. The orthorhombic 
phase is stable only at the single critical concentration, $x_{c}$, which 
just corresponds to MPB point.
}
\label{fig:4}
\end{figure}

More generally, the free energy $F_{el}(\rho)$ and $F_{el}(\rho')$ with 
an arbitrary polarization direction is given by,
\begin{eqnarray}
\ F_{el}(\rho) = -\frac{g'^{2}}{12c'}(3\rho-1)^{2} 
\ \ \ \ \ \ \ \ \ \ \  \ \ \ \ \ \nonumber \\
- \frac{g''^{2}}{8c''}(1-\rho)(1+3\rho),   \label{eq:16}
\end{eqnarray} 
\begin{eqnarray}
\ F_{el}(\rho') = -\frac{g'^{2}}{12c'} 
- \frac{g'^{2}}{4c'}(1-2\rho')^{2}    \ \  \ \ \ \ \ \    \nonumber \\
- \frac{g''^{2}}{2c''} \rho'(1-\rho').   \label{eq:17}
\end{eqnarray}                               
It is easily seen that at MPB where eq. (\ref{eq:15}) is satisfied,
\begin{eqnarray}
\ F_{el}(\rho) = \ F_{el}(\rho') = A,  \label{eq:18}
\end{eqnarray}
irrespective of the values of $\rho$ and $\rho'$ which means that at MPB, 
the energy is completely degenerated throughout the postulated path of 
polarization rotation. Only by the introduction of higher order energy 
terms which are neglected in the present treatment, the degeneracy will 
be lifted so that either of O-, $\mathrm{M_{a}}$- or $\mathrm{M_{c}}$- phases 
may become stabilized in the MPB region as shown 
by Vanderbilt and Cohen.\cite{rf:10}
This situation may explain the 
experimental observations that depending on subtle differences of the 
prehistory of sample preparation, the symmetry of the stable state 
becomes different.

Furthermore, neglecting the higher order terms, we see that at $x=x_{c}$ 
the system could move over 
between R$\Leftrightarrow$$\mathrm{M_{a}}$$\Leftrightarrow$T$\Leftrightarrow$$\mathrm{M_{c}}$$\Leftrightarrow$O 
under external forces such as electric field and stress without 
any cost of elastic energy. Therefore, at MPB some of the 
susceptibility tensor components of $\chi_{ij}$ (dielectric constants), 
$d_{ijk}$ (piezoelectric constants) and the inverse of $c_{ijkl}$ 
(elastic constants) should critically blow up, although in 
the real system, the higher order effects would tend to prevent 
such a strong anomaly.
Actually, Ishibashi and Iwata\cite{rf:m1,rf:m2,rf:m3} have given the 
explicit expressions for these susceptibility tensors, some of which
tend to show the critical behavior.
The above simple energy considerations seem 
to reveal the essential origin of the anomalous physical properties at MPB.

As a further application of this view point, we take notice on the 
elastic anomaly exhibited by hexagonal $\mathrm{BaTiO_{3}}$.\cite{rf:11}
It is known that hexagonal $\mathrm{BaTiO_{3}}$ (h-BT) undergoes two 
phase transitions successively as hexagonal $\rightarrow$ orthorhombic (space 
group $\mathrm{C222_{1}}$) $\rightarrow$ monoclinic (space group 
$\mathrm{P2_{1}}$).\cite{rf:12,rf:13} The relevant order parameter is the 
doubly degenerated phonon modes, $\mathrm{Q_{1}}$ and $\mathrm{Q_{2}}$. 
Ishibashi\cite{rf:14,rf:15} pointed out that the transition 
between O- and M- phases is caused by the rotation of the polarization 
vector of the mode in the 
2-dimensional ($\mathrm{Q_{1}}$,$\mathrm{Q_{2}}$)-space, which 
is very similar to the feature presented in this paper if we 
replace $\mbox{\boldmath$P$}$ by $\mbox{\boldmath$Q$}$. 
However, it should be noticed 
that the anomaly at the O-M transition region in h-BT is manifested 
in the elastic compliances rather than piezoelectricity because 
the Q-modes are optically inactive. In this context, h-BT may be 
considered to provide another example belonging to the different 
category concerning MPB adventure.

\section*{Acknowledgements}

Financial supports of Grant-In-Aid for Science Research from 
Monbu-Kagakusho, Grant for Development of New Technology from 
Shigaku-Shinkozaidan , Waseda University Grant for Special 
Research Projects and US DOE under contract No.DE-AC0298CH10866 
are also gratefully acknowledged. This work was performed under 
US-Japan Cooperative Neutron Research Program.

\end{document}